\documentclass[aps,twocolumn,showpacs,superscriptaddress]{revtex4}

\usepackage[dvips]{graphicx}
\usepackage{amsmath}
\usepackage{amsfonts}
\usepackage{amssymb} 
\newcommand{\myfig}[4]
{
\begin{figure*}
\begin{tabular}{@{}cc@{}}
\resizebox{0.35\textwidth}{!}{%
\includegraphics{./#1.eps}
}
&
\resizebox{0.35\textwidth}{!}{%
\includegraphics{./#3.eps}
}
\\
\stepcounter{figure}\label{#1}
\begin{minipage}[c]{0.49\textwidth}
\footnotesize \emph{Fig. \thefigure.} #2
\end{minipage}
&
\stepcounter{figure}\label{#3}
\begin{minipage}[c]{0.49\textwidth}
\footnotesize \emph{Fig. \thefigure.} #4
\end{minipage}
\end{tabular}
\end{figure*}
}

\begin{document}
\title{Nanotube Quantum Dot Transport With Spin-Orbit Coupling and Interacting Leads}
\author{Ogloblya O.V. }
\author{Kuznetsova G.M. }
\affiliation{Taras Shevchenko National University, 64/13 Volodymyrska St., Kyiv 01601, Ukraine }
\date{\today}
\begin{abstract}
To date the impact of spin bias on the Kondo effect in nanotube quantum dots (QDs) is scarcely explored, especially in the case of significant Coulomb interaction between quantum dots and electric contact leads.  Recent rapid experimental progress in nanotechnology opens new possibilities to study spin-bias-induced transport phenomena. Thus, of great interest are theoretical investigations of these transport properties in comparison with the case of a conventionally applied voltage for nanotube QDs interacting with contact leads.  Such an investigation was carried out in this work where we analyzed the effects of a spin voltage as well as a conventionally applied voltage in a QD 
system with a different number of quantum states in the dot region in presence of Coulombic interaction between the quantum dot and two leads.  
The transport is described within the Keldysh non-equilibrium Green's function (NEGF) framework. 
We extended the NEGF treatment developed for noninteracting leads onto the case of four quantum states 
m = $ \{\sigma , \lambda \}  =  \{\pm , \pm\}$  ( two spins and two energy subbands,
 which is the case for a nanotube QD) interacting with leads.
  Our derivation is based on the equation-of-motion technique and Langreth's theorem.  For a Coulombic repulsion between the contacts and QD we obtain an expression for the current through QD for the four quantum states. 
To determine the parameters of the model Hamiltonian we used our previous calculations~\cite{one} of the electronic properties of a symmetrical nanotube QD (5,5)/(10,0)\_1/(5,5)
in a tight binding model, where \_1 denotes the length of the middle QD segment of a (10,0) zigzag nanotube.  
We calculated the density of electronic states with spin up and down for the case of a single QD without pseudospin states for an infinite Coulomb repulsion, in good agreement with the calculations of Yuan Li, et al.~\cite{two}.  Our calculation showed that the position of the conductance peaks nearest to zero is not affected by the strength of the QD-lead Coulombic interaction parameters. 
 We also demonstrated that this interaction shifts the density of states to higher energies.  The interplay between the Kondo effect and the bias is highly temperature-dependent and becomes significant only at low temperatures. Lastly, we found that the existence of four quantum states
  $m = \{\sigma, \lambda\} = \{\pm, \pm\}$ 
leads to abrupt changes in the density of states. In this case the values of the current are approximately ten times lower than for QD with only two quantum 
states  $m = \{\sigma \} = \{ \pm \}$.  However, in the case of a conventional bias the current amplitudes in both cases are approximately the same.
\end{abstract}

\maketitle

\section{Introduction}

A fascinating many-body phenomenon in condensed matter is Kondo effect. It was extensively studied during the past years in connection to the increase of  applications of carbon nanotubes as quantum dot (QD) in variety nanoelectronic devices. Fabrications of carbon nanotubes opens new stage in the investigation of Kondo effect~\cite{three}. In carbon nanotubes there is additional orbital degrees of freedom which originates from two electronic subbands near Fermi energy and plays a role of pseudospin.
Recently, the results by Kuemmeth et al.~\cite{kuem} demonstrate that the spin and orbital motion of electrons are strongly coupled, thereby breaking the SU(4) symmetry of electronic states in a such system. The spin-orbit coupling determines the filling order in the two-electron ground state~\cite{kuem} and leads to the fact that Kondo effects manifests as split resonant peaks in the differential conductance even at zero magnetic field. In recent years, experimental progress opens new possibility for the study of spin-bias-induced transport in mesoscopic systems. For example spin bias can be achieved by controlling spin accumulation at the biased contacts between ferromagnetic and nonmagnetic leads. This allows one to generate a pure spin current without any accompanying charge current~\cite{four,five}. However the effect of the spin bias on the Kondo effect remains weakly explored, especially in situation where the Coulomb interaction of QD with leads are significant. More importantly, the Kondo effect in nanotube QD in the presence of  Coulomb interactions between the QD and two adjacent leads has not been studied.
 The need of extending steady state transport formalism is required because new experimental results can't  be explained with steady state transport formalism.
 Our goal was to derive such a formalism that could fit CN QD and can account for dot-lead electron interraction.
\section{Model and Formulation}
We investigate model system with four-state QD  with interplay between the spin and orbital (pseudospin) degrees of freedom. This situation highly relevant to CN QD and  is more general than two-state QD investigated previously (see for example ~\cite{two}). We use the same strategy as in work ~\cite{two} but the index that enumerates quantum dot state can run over four possible quantities that lead sums and some another terms in formulas. Therefore this goal requires complete rederivation of formulas of ~\cite{two}. Of course the formulas of ~\cite{two} should be acquired as partial case from a more general expression. 

Let's use Keldysh formalism available for the case of one quantum dot with two energy levels split by spin-orbit coupling of electrons with different spin states~\cite{two} and let's make it acceptable to the case 
of symmetrically connected carbon nanotubes (CNT) where there is additional orbital degrees of freedom which originates from two electronic subbands near Fermi energy. Such a degree of freedom plays the role of pseudospin. For the nanotube quantum dot we adopted the following Hamiltonian $H=H_d+H_c+H_T+H_{cd}$. The first term is the Hamiltonian which models an isolated CNT QD 
\begin{align}
H_d=\sum_m \varepsilon_m d_m^{+}d_m+\frac{U}{2} \sum_{m, m^\prime}^{m\neq m^\prime} n_m n_{m^\prime}\;,
\end{align}
where $n_{m}=d_m^{+}d_m$
and 
$d_{m}^{+}$  creates, $d_{m}$ 
 annihilates electron in the dot with $m=\{\sigma,\lambda\}$ configuration, where  
$\{\sigma,\lambda\}=\{\pm,\pm\}$ are the spin and orbital quantum numbers,  $U$ is the on-site Coulomb repulsion,  $\varepsilon_m=\varepsilon_d-\sigma \lambda \Delta_{SO}/2$ is the single-particle energy, its first term $\varepsilon_d$ is the basic dot level and the second accounts for spin-orbit coupling with the constant of a such interaction 
$\Delta_{SO}$.  $H_c=\sum_{\alpha k m} \epsilon_{\alpha k m} c_{\alpha k m}^{+}c_{\alpha k m}$ - Hamiltonian of the contacts,  $\alpha=L,R$ - 
enumerates the left lead (L) and right (R) lead respectively,
  $c_{\alpha k m}^{+}$ and $c_{\alpha k m}$ creates and annihilates electron with wave vector $k$ and spin-pseudospin state $m$ in lead   $\alpha=L,R$. The tunneling between the dot and the leads is described  by  $H_T=\sum_{\alpha k m} t_{\alpha m} c_{\alpha k m}^{+}d_m+H.c.$ 
with spin and orbital conservation since the length of middle nanotube is very short (one translational period). 
In the case of conventional nanotube QD 
which is composed from one single walled carbon nanotube deposited on substrate
to which gate voltage is applyed the role of the leads plays the same nanotube 
thus leads have the same orbital symmetry~\cite{six,seven}.
Therefore for conventional NT QD spin and orbital conservation are valid.
Our approach works for both cases conventional QD and QD from heterojunctions $(n,n)/(2n,0)\_1/(n,n)$ but 
for obtaining numerical results we choose parameters for QD from heterojunctions $(n,n)/(2n,0)\_1/(n,n)$.
We also incorporate the Coulombic interaction between the charges on the quantum dot and two contacts. This interaction could play significant role especially due to the proximity of the contacts to dot. 
This last term in the Hamiltonian is given by~\cite{eight}:
\begin{align}
H_{cd}=\sum_{\alpha,k,m,m^\prime} I_{\alpha k} d_m^{+}d_m c_{\alpha k m^\prime}^{+}c_{\alpha k m^\prime}\;.
\end{align}

Firstly we obtained a general formula of the current through CNT QD by considering the interaction between the QD and contacts. The current from the lead  $\alpha$ through the barrier to the dot can be calculated from the time evolution of the occupation number operator of the lead $\alpha$: $J_{\alpha m}=$ $-e \langle\dot{N}_{\alpha m}\rangle=$ $-\frac{ie}{\hbar}\langle [H,N_{\alpha m}]\rangle$  where 
$N_{\alpha m}=$ $\sum_k c_{\alpha k m}^{+}c_{\alpha k m}$ . Evaluating commutators we get
\begin{align}
J_{\alpha m}=\frac{ie}{\hbar}\biggl( t_{\alpha m}\langle c_{\alpha k m}d_m\rangle-t_{\alpha m}^{*}\langle d_m^{+}c_{\alpha k m}\rangle \biggr) \;.
\end{align}
By using the definition of less  Green's function $G^{<}_{\alpha k m} =\\ i \langle c_{\alpha k m}^{+}(t^\prime)d_m(t)\rangle$,
the current can be written as 
\begin{align}
J_{\alpha m}=\frac{2e}{\hbar}Re[\sum_k t_{\alpha m} G^{<}_{\alpha k m}(t,t)] \;.
\end{align}
Let's consider the equation of motion for the time-ordered Green's function $G^{t}_{\alpha k m}(t,t^\prime)$
\begin{flalign}
-i\hbar \frac{\partial}{\partial t^\prime}G^{t}_{\alpha k m}(t,t^\prime)=\epsilon_{\alpha k m} G^{t}_{\alpha k m}(t,t^\prime)+ 
t_{\alpha m}^{*} G^{t}_m(t,t^\prime)-\nonumber\\
i I_{\alpha k} \sum_{m^\prime} T\{ d_m(t)d_{m^\prime}^{+}(t^\prime)d_{m^\prime}(t^\prime) c_{\alpha k m}^{+}(t^\prime)\}
\end{flalign}
where we use the definition of time-ordered Greeen's function of dot $G^{t}_m(t,t^\prime)=$ $-i\langle T\{d_m(t)d^{+}_m(t^\prime)\}\rangle$,
here T is the  time ordering operator. The last term prevents obtaining a closed form for
$G^{t}_{\alpha k m}(t,t^\prime)$. At this step we need to use some approximations. The simplest possible solution is to use Hartree-Fock approximation
\begin{align}
-i\langle T\{d_m(t)d_{m^\prime}^{+}(t^\prime) d_{m^\prime}(t^\prime) c_{\alpha k m}^{+}(t^\prime)\}\rangle\approx \nonumber \\
n_{m^\prime}(-i)\langle T\{d_m(t)c_{\alpha k m}^{+}(t^\prime)\}\rangle=
n_{m^\prime}G_{\alpha,k,m}^{t}(t,t^\prime)\nonumber
\end{align}
where $n_{m^\prime}=\langle d_{m^\prime}^{+}d_{m^\prime}\rangle$. Therefore in Hartree-Fock approximation the equation of motion for the time-ordered Green's function is:
$(-i \hbar \frac{\partial }{\partial t^\prime} -\epsilon^\prime_{\alpha k m})G_{\alpha k m}^t(t,t^\prime)=\\ t_{\alpha m}^{*}G_m^t(t,t^\prime)$,
where $\epsilon_{\alpha k m}^\prime=$ $\epsilon_{\alpha k m}+A_{\alpha k}$, $A_{\alpha k}=$ $n I_{\alpha k}$, $n=$ $\sum_m n_m$. 
If we denote $g_{\alpha k m}^t(t,t^\prime)$ Green's function for operator $(-i \hbar \frac{\partial }{\partial t^\prime} -\epsilon^\prime_{\alpha k m})$ and operating them to previous equation from right we get $G_{\alpha k m}^t(t,t^\prime)=\\ \int dt_1 t_{\alpha m}^{*} G_m^t(t,t_1) g_{\alpha k m}^t(t_1,t^\prime)$.
Similarly the contour-ordered Green's function can be written as
$G_{\alpha k m}(\tau,\tau^\prime)=\\ \int d\tau_1 t_{\alpha m}^{*}G_m(\tau,\tau_1)g_{\alpha k m}(\tau_1,\tau^\prime)$. 
Now using Langreth's theorem we can get $G_{\alpha k m}^{<}(t,t^\prime)$
\begin{align}
G_{\alpha k m}^{<}(t,t^\prime)=\int d\tau_1 t_{\alpha m}^{*}G_m^r(\tau,\tau_1)g_{\alpha k m}^{<}(\tau_1,\tau^\prime)+
\nonumber \\
\int d\tau_1 t_{\alpha m}^{*}G_m^{<}(\tau,\tau_1)g_{\alpha k m}^a(\tau_1,\tau^\prime)
\end{align}
where the Green's function of leads for uncoupled system with modified energy levels $\epsilon_{\alpha k m}^\prime$ is:
$g_{\alpha k m}^{<}(t,t^\prime)=\\ i f_{\alpha m}(\epsilon_{\alpha k m}^\prime)e^{-i\epsilon_{\alpha k m}^\prime(t-t^\prime)}$,
$g_{\alpha k m}^a(t,t^\prime)=$ $i \theta (t^\prime-t) e^{-i \epsilon_{\alpha k m}^\prime(t-t^\prime)}$
and  $f_{\alpha m}(\epsilon_{\alpha k m}^\prime)=\bigl( e^{\frac{\epsilon_{\alpha k m}^\prime-\mu_{\alpha m}}{k T}}+1\bigr)^{-1}$ is the Fermi distribution.
After substitution of previous formulas to (6) and then to (4) we get  
\begin{flalign}
J_{\alpha m}=\frac {ie}{2\pi\hbar}\int dz\Gamma_{\alpha m}(z) (G_m^{<}(z+A_{\alpha}(z))+\nonumber\\
f_{\alpha m}(z+A_{\alpha}(z)G_m^r(z+A_{\alpha}(z))-\nonumber\\
f_{\alpha m}(z+A_{\alpha}(z)G_m^a(z+A_{\alpha}(z)))
\; , \nonumber
\end{flalign}
where $\Gamma_{\alpha m}(z)=2\pi \sum_k |t_{\alpha m}|^2 \delta(z-\epsilon_{\alpha k m})$ is the coupling function. We will consider proportional couplings to the leads
$\Gamma_{Lm}=\chi \Gamma_{Rm}$  and symmetrize this formula by calculating the current $J=x J_L-(1-x)J_R$  where $J_\alpha=\sum_m J_{\alpha m}$  and where the parameter $x$  can be chosen as
$x=1/(1+\chi)$.
For the case of proportional couplings we get:
\begin{align}
J=\frac{e}{\hbar}\sum_m \int dz \Gamma_m(z) (f_{Lm}(z+A_L(z))\rho_m(z+A_L(z))-\nonumber\\
f_{Rm}(z+A_R(z))\rho_m(z+A_R(z))\; ,
 \end{align}
where $\Gamma_m(z)=\frac{\Gamma_{Lm}(z)\Gamma_{Rm}(z)}{\Gamma_{Lm}(z)+\Gamma_{Rm}(z)}$, $\rho_m(z)=-\frac{1}{\pi}Im[G_m^r(z)]$. Using last formula only demands to know
$G_m^r(z)$.  With the goal of finding retarded Green's function $G_m^r(z)$ we use standard equation of motion technique based on the general relation
\begin{align}
\ll\hat{F}_1|\hat{F}_2\gg^r z=\langle\{\hat{F}_1|\hat{F}_2\}\rangle+\ll [\hat{F}_1,\hat{H}]|\hat{F}_2\gg^r\nonumber\; .
 \end{align}
Writing this relation for the case $\hat{F}_1=d_m$ and $\hat{F}_2=d_m^{+}$  we get 
\begin{align}
\ll d_p|d_p^{+}\gg(z-\varepsilon_p)=1+U\sum_q^{q\neq p} \ll d_q^{+} d_q d_p|d_p^{+}\gg + \nonumber\\
\sum_{\alpha k} t_{\alpha p}^{*}\ll c_{\alpha k p}|d_p^{+}\gg 
-2\sum_{\alpha k} t_{\alpha p}^{*}
\ll c_{\alpha k p} d_p^{+}d_p | d_p^{+}\gg +\nonumber\\
\sum_{\alpha k q} I_{\alpha k}\ll c_{\alpha k q}^{+}c_{\alpha k q}d_p|d_p^{+}\gg  \; . 
\end{align}
In this formula and below under by $\ll \hat{F}_1,\hat{F}_2\gg $  we understand $\ll \hat{F}_1,\hat{F}_2\gg ^r$  for the sake of simpli?cation in writing formulas. For the dot region we use indexes
$p,q,r$ and $\alpha , \beta$  for numbering the left $\alpha=L$, $\beta=L$ and right  $\alpha=R$, $\beta=R$ contacts. For wave vector in leads we use $k$,$\tilde{k}$ . 
In the right hand side of expression (8) appears new Green's function for each of which we also can write equations of motion 
\begin{align}
\ll c_{\alpha k p}|d_p^{+}\gg (z-\epsilon_{\alpha k p})=t_{\alpha p}\ll d_p|d_p^{+}\gg - \nonumber \\
2 t_{\alpha p}\ll c_{\alpha k p}^{+}c_{\alpha k p} d_p|d_P^{+}\gg 
+I_{\alpha k}\sum_q c_{\alpha k p}d_q^{+}d_q|d_p^{+}\gg  \; . \nonumber
\end{align}
In the last formula we neglect three-operator terms because they contain high order spin correlation. We do the same  for all other formulas. Whenever a term with $\hat{F}_1$  contains 
five or more operators in
$\ll \hat{F}_1|d_p^{+}\gg $   we drop it. For  $p\neq q$ we can write:
\begin{align}
\ll d_q^{+}d_q d_p | d_p^{+}\gg (z-\varepsilon_p-U)=n_q+ \nonumber \\
\sum_{ak} t_{\alpha p}^{*}\ll c_{\alpha k p}d_q^{+}d_q|d_p^{+}\gg +
\sum_{\alpha k} t_{\alpha q}^{*}\ll c_{\alpha k q}d_q^{+}d_p|d_p^{+}\gg -\nonumber\\
\sum_{\alpha k} t_{\alpha q}\ll c_{\alpha k q}^{+}d_q d_p|d_p^{+}\gg \; ,
\end{align}
\begin{align}
\ll c_{\alpha k p}d_q^{+}d_q| d_p^{+}\gg (z-\epsilon_{\alpha k p}-I_{\alpha k})= \nonumber\\
t_{\alpha p}\ll d_q^{+}d_q d_p|d_p^{+}\gg +
\sum_{\beta k^\prime} t_{\beta q}^{*}\ll c_{\alpha k p} c_{\beta k^\prime q}d_q^{+}|d_p^{+}\gg -\nonumber\\
\sum_{\beta k^\prime} t_{\beta q}\ll  c_{\beta k^\prime q}^{+} c_{\alpha k p}d_p|d_p^{+}\gg \; ,
\end{align}
\begin{align}
\ll c_{\alpha k p}d_p^{+}d_p| d_p^{+}\gg (z-\epsilon_{\alpha k p}-I_{\alpha k})=\nonumber\\
\sum_{\beta k^\prime} t_{\beta p}^{*}\ll c_{\beta k^\prime p} c_{\alpha k p} d_p^{+}|d_p^{+}\gg -\nonumber\\
\sum_{\beta k^\prime} t_{\beta p}\ll  c_{\beta k^\prime p}^{+} c_{\alpha k p}d_p|d_p^{+}\gg \; ,
\end{align}
\begin{align}
\ll c_{\alpha k q}d_q^{+}d_p| d_p^{+}\gg (z-\varepsilon_p+\varepsilon_q-\epsilon_{\alpha k p}-I_{\alpha k})=
\langle c_{\alpha k q} d_q^{+}\rangle+\nonumber\\
t_{\alpha q}\ll d_q^{+}d_q d_p|d_p^{+}\gg 
+\sum_{\beta k^\prime} t_{\beta p}^{*}\ll c_{\alpha k q} c_{\beta k^\prime p}  d_q^{+}|d_p^{+}\gg -\nonumber\\
\sum_{\beta k^\prime} t_{\beta q}\ll  c_{\beta k^\prime q}^{+} c_{\alpha k q}d_p|d_p^{+}\gg \; ,
\end{align}
\begin{align}
\ll c_{\alpha k q}^{+}d_q d_p| d_p^{+}\gg (z-U-\varepsilon_p-\varepsilon_q+\epsilon_{\alpha k p})=\nonumber\\
\sum_{\beta k^\prime} t_{\beta q}^{*}\ll c_{\alpha k q}^{+} c_{\beta k^\prime q}  d_p|d_p^{+}\gg +\nonumber\\
\sum_{\beta k^\prime} t_{\beta p}^{*}\ll  c_{\alpha k q}^{+}c_{\beta k^\prime p} d_q|d_p^{+}\gg -t_{\alpha q}^{*}\ll d_q^{+}d_q d_p|d_p^{+}\gg \; ,
\end{align}
\begin{align}
\ll c_{\alpha k q}^{+}d_q^{+} d_p| d_p^{+}\gg (z-\varepsilon_p+\varepsilon_q+\epsilon_{\alpha k q}+I_{\alpha k})=\nonumber\\
\sum_{\beta k^\prime} t_{\beta p}^{*}\ll c_{\alpha k q}^{+} c_{\beta k^\prime p}  d_q^{+}|d_p^{+}\gg -\nonumber\\
\sum_{\beta k^\prime} t_{\beta q}\ll  c_{\beta k^\prime q}^{+} c_{\alpha k q}^{+} d_p|d_p^{+}\gg \;.
\end{align}
We used decoupling approximation based on the following rules~\cite{nine} $\langle Y X\rangle=0$, $\ll Y X_1 X_2|d_p^{+}\gg \approx \langle X_1 X_2\rangle \\ \ll Y|d_p^{+}\gg$ where $X$ 
represents an operator of contacts and $Y$  an operator of quantum dot.
\begin{align}
\ll d_p| d_p^{+}\gg (z-\varepsilon_p-\sum_{\alpha k q}I_{\alpha k}\langle c_{\alpha k q}^{+}c_{\alpha k q}\rangle)=
\nonumber\\
1+U\sum_q^{q\neq p} \ll d_q^{+}d_q d_p|d_p^{+}\gg +\nonumber\\
\sum_{\alpha k} t_{\alpha p}^{*}\ll  c_{\alpha k p}|d_p^{+}\gg -
2 \sum_{\alpha k} t_{\alpha p}^{*}\ll  c_{\alpha k p}d_p^{+}d_p|d_p^{+}\gg \;,
\end{align}
\begin{align}
\ll c_{\alpha k p}| d_p^{+}\gg (z-\epsilon_{\alpha k p})=t_{\alpha p}(1-2n_{\alpha k p})\ll d_p|d_p^{+}\gg \nonumber\\
+I_{\alpha k p}\ll c_{\alpha k p} d_p^{+}d_p|d_p^{+}\gg +I_{\alpha k}\sum_q^{q\neq p} \ll  c_{\alpha k p}d_q^{+}d_q|d_p^{+}\gg \;,
\end{align}
\begin{align}
\ll c_{\alpha k p} d_p^{+}d_p|d_p^{+}\gg (z-\epsilon_{\alpha k p}-I_{\alpha k})= \nonumber\\
-t_{\alpha p} n_{\alpha k p}\ll d_p|d_p^{+}\gg \;,
\end{align}
\begin{align}
\ll d_q^{+}d_q d_p| d_p^{+}\gg (z-\varepsilon_p-U)=n_q +\nonumber\\
\sum_{\alpha k} \bigl( t_{\alpha p}^{*}\ll c_{\alpha k p}d_q^{+}d_q|d_p^{+}\gg  +\nonumber\\
t_{\alpha q}^{*}\ll c_{\alpha k q}d_q^{+}d_p|d_p^{+}\gg -t_{\alpha q}\ll c_{\alpha k q}^{+}d_q d_p|d_p^{+}\gg \bigr) \;,
\end{align}
\begin{align}
\ll c_{\alpha k p}d_q^{+}d_q| d_p^{+}\gg (z-\epsilon_{\alpha k p}-I_{\alpha k})=\nonumber\\
t_{\alpha p} \ll d_q^{+}d_q d_p|d_p^{+}\gg \;,
\end{align}
\begin{align}
\ll c_{\alpha k p}d_q^{+}d_p| d_p^{+}\gg (z-\epsilon_p+\varepsilon_q-\epsilon_{\alpha k q}-I_{\alpha k})=\nonumber\\
t_{\alpha q} \ll d_q^{+}d_q d_p|d_p^{+}\gg -t_{\alpha q} n_{\alpha k q}\ll d_p|d_p^{+}\gg \;,
\end{align}
\begin{align}
\ll c_{\alpha k q}^{+}d_q d_p| d_p^{+}\gg (z-U-\varepsilon_p-\varepsilon_q+\epsilon_{\alpha k q})=\nonumber\\
t_{\alpha q}^{*} n_{\alpha k q}\ll d_p|d_p^{+}\gg -t_{\alpha q}^{*}\ll d_q^{+} d_q d_p|d_p^{+}\gg \;.
\end{align}
From (16) using (18, 19) we get:
\begin{align}
\ll c_{\alpha k p}| d_p^{+}\gg =\frac{t_{\alpha p}(1-2n_{\alpha k p})}{z-\epsilon_{\alpha k p}}\ll d_p|d_p^{+}\gg-
\nonumber\\ \frac{I_{\alpha k}t_{\alpha p} }{(z-\epsilon_{\alpha k p})(z-\epsilon_{\alpha k p}-
I_{\alpha k})}
\nonumber\\
\bigl(n_{\alpha k p}\ll d_p|d_p^{+}\gg +\sum_q^{q\neq p}\ll d_q^{+}d_q d_p|d_p^{+}\gg \bigr)\;.\nonumber
\end{align}
The other formulas are obtained in a similar way:
\begin{align}
\ll d_p| d_p^{+}\gg \biggl(z-\varepsilon_p-\sum_{\alpha k q} I_{\alpha k}\langle c_{\alpha k q}^{+}c_{\alpha k q}\rangle-
\nonumber\\
\sum_{\alpha k}
\frac{|t_{\alpha p}|^2 }{z-\epsilon_{\alpha k p}}\bigl(1+\frac{I_{\alpha k} n_{\alpha k p}}{z-\epsilon_{\alpha k p}-I_{\alpha k}}\bigr)\biggr)=1+\nonumber\\
\bigl(U+\sum_{\alpha k} \frac{|t_{\alpha p}|^2 I_{\alpha k}}{(z-\epsilon_{\alpha k p})(z-\epsilon_{\alpha k p}-I_{\alpha k})}\bigr)\nonumber\\
\sum_q^{q \neq p} \ll d_q^{+}d_q d_p|d_p^{+}\gg \;.\nonumber
\end{align}
After introduction
\begin{align}
\Sigma_p^{(0)}(z)=\sum_{\alpha k} \frac{|t_{\alpha p}|^2}{z-\epsilon_{\alpha k p}}\;,\nonumber\\
\Sigma_p^{(1)}(z)=\sum_{\alpha k} \frac{|t_{\alpha p}|^2 I_{\alpha k}}{(z-\epsilon_{\alpha k p})(z-\epsilon_{\alpha k p}-I_{\alpha k})}\;,\nonumber \\
\tilde{\Sigma}_p^{(1)}(z)=\sum_{\alpha k} \frac{|t_{\alpha p}|^2 I_{\alpha k} n_{\alpha k p}}{(z-\epsilon_{\alpha k p})(z-\epsilon_{\alpha k p}-I_{\alpha k})}\;,\nonumber\\
\Sigma_p^{(2)}=\sum_{\alpha k q}  I_{\alpha k}n_{\alpha k q}\;,
 \end{align}
we get:
\begin{align}
\ll d_p d_p^{+}\gg \bigl(z-\varepsilon_p-\Sigma^{(2)}-\Sigma_p^{(0)}(z)-\tilde{\Sigma}_p^{(1)}(z)\bigr)=\nonumber\\
1+\bigl(U+\Sigma_p^{(1)}(z)\bigr)\sum_q^{q \neq p}\ll d_q^{+}d_q d_p| d_p^{+}\gg \;.
 \end{align}
From (18) using (19,20,21) in a similar way we get:
\begin{align}
\ll d_q^{+}d_qd_p|d_p^{+}\gg \biggl( z-\varepsilon_p-U-\sum_{\alpha k}
\frac{|t_{\alpha p}|^2}{z-\epsilon_{\alpha k p}-I_{\alpha k}}
\nonumber\\
-\sum_{\alpha k}\bigl( 
\frac{|t_{\alpha q}|^2}{z-\varepsilon_p+\varepsilon_q-\epsilon_{\alpha k q}-I_{\alpha k}}
\nonumber\\
+\frac{|t_{\alpha q}|^2}{z-U-\varepsilon_p-\varepsilon_q+\epsilon_{\alpha k q}}\bigr) \biggr)=
\nonumber\\
n_q-\ll d_p|d_p^{+}\gg \sum_{\alpha k}\bigl( 
\frac{|t_{\alpha q}|^2 n_{\alpha k q}}{z-\varepsilon_p+\varepsilon_q-\epsilon_{\alpha k q}-
I_{\alpha k}}+
\nonumber\\
\frac{|t_{\alpha q}|^2 n_{\alpha k q}}{z-U-\varepsilon_p-\varepsilon_q+\epsilon_{\alpha k q}}\bigr)\;,\nonumber
\end{align}
After introducing:
\begin{align}
A_{qp}(z)=\sum_{\alpha k}\bigl( \frac{|t_{\alpha q}|^2}{z-\varepsilon_p+\varepsilon_q-\epsilon_{\alpha k q}-I_{\alpha k}}+
\nonumber\\
\frac{|t_{\alpha q}|^2}{z-U-\varepsilon_p-\varepsilon_q+\epsilon_{\alpha k q}}\bigr)\;,\nonumber\\
\tilde{A}_{qp}(z)=\sum_{\alpha k}\bigl( \frac{|t_{\alpha q}|^2 n_{\alpha k q}}{z-\varepsilon_p+\varepsilon_q-\epsilon_{\alpha k q}-I_{\alpha k}}+
\nonumber\\
\frac{|t_{\alpha q}|^2 n_{\alpha k q}}{z-U-\varepsilon_p-\varepsilon_q+\epsilon_{\alpha k q}}\bigr)\;,
 \end{align}
we can rewrite it as:
\begin{align}
\ll d_q^{+}d_q d_p|d_p^{+}\gg \bigl(z-\varepsilon_p-U-\sum_{\alpha k}\frac {|t_{\alpha p}|^2}{z-\epsilon_{\alpha k p}-I_{\alpha k}}-\nonumber\\
A_{qp}(z)\bigr)=
n_q-\ll d_p|d_p^{+}\gg \tilde{A}_{qp}(z)\;.
 \end{align}
The last formula (25) allows as to write (23) in the form: 
\begin{align}
\ll d_p|d_p^{+}\gg \biggl(z-\varepsilon_p-\Sigma_p^{(0)}(z)-\tilde{\Sigma}_p^{(1)}(z)-\Sigma^{(2)}+
\nonumber\\
\bigl(U+\Sigma_p^{(1)}(z)\bigr)\sum_q^{q \neq p} \frac{\tilde{A}_{qp}(z)}{z-\varepsilon_p-U-\tilde{\Sigma}_p^{(0)}(z)-A_{qp}(z)}\biggr)\nonumber\\
=1+\bigl(U+\Sigma_p^{(1)}(z)\bigr)\sum_q^{q \neq p} \frac{n_q}{z-\varepsilon_p-U-\tilde{\Sigma}_p^{(0)}(z)-A_{qp}(z)}\;.
 \end{align}
where
$\tilde{\Sigma}_p^{(0)}(z)=\sum_{\alpha k}\frac{|t_{\alpha p}|^2}{z-\epsilon_{\alpha k p}-I_{\alpha k}}$.  The last equation (26) can be solved for $\ll d_p|d_p^{+}\gg =G_p(z)$ :
\begin{align}
G_p(z)=\biggl(1+\nonumber\\
\bigl(U+\Sigma_p^{(1)}(z)\bigr)
\sum_q^{q \neq p}\frac{n_q}{z-\varepsilon_p-U-\tilde{\Sigma}_p^{(0)}(z)-A_{qp}(z)}\biggr)\nonumber\\
/\biggl(
z-\varepsilon_p-\Sigma_p^{(0)}(z)-\tilde{\Sigma}_p^{(1)}(z)-\Sigma^{(2)}+\nonumber\\
\bigl(U+\Sigma_p^{(1)}(z)\bigr)
\sum_q^{q \neq p} \frac{\tilde{A}_{qp}(z)}{z-\varepsilon_p-U-\tilde{\Sigma}_p^{(0)}(z)-A_{qp}(z)}\biggr)
\end{align}
here $n_{\alpha k p}=f_{\alpha p}(\epsilon_{\alpha k p})$,
$f_{\alpha p}(\epsilon_{\alpha k p})=\bigl(exp(\frac{\epsilon_{\alpha k p}-\mu_{\alpha p}}{k T})+1\bigr)^{-1}$
- Fermi distribution, $\mu_{\alpha p}$ - chemical potential of lead $\alpha$ for spin and orbital quantum number $p=\{\sigma,\lambda\}$. 

Let us explore the limit $U\rightarrow \infty$. By taking this limit in (27) we get
\begin{align}
G_p(z)=\frac{1-\sum_q^{q\neq p} n_q}{z-\varepsilon_p -\Sigma_p^{(0)}(z)-\tilde{\Sigma}_p^{(1)}(z)-\Sigma^{(2)}-\Sigma_p^{(3)}(z)}\;,
 \end{align}
where $\Sigma_p^{(3)}(z)=\sum_q^{q\neq p}\sum_{\alpha k}\frac{|t_{\alpha q}|^2 n_{\alpha k q}}{z-\varepsilon_p+\varepsilon_q-\epsilon_{\alpha k q}-I_{\alpha k}}$. 
The Green's function 
$G_p(z)$ depends on unknown quantities $n_q=\langle d_q^{+}d_q\rangle$. But $n_q=-\frac{i}{2\pi}\int dz G_q^{<}(z)$ and using Langreth's theorem we get  
\begin{align}
n_q=-\frac{1}{\pi}\int dz \frac{\Gamma_{Lq}(z)f_{Lq}(z)+\Gamma_{Rq}(z)f_{Rq}(z)}{\Gamma_{Lq}(z)+\Gamma_{Rq}(z)} Im \bigl[ G_q^r(z) \bigr]\;.
\end{align}
Formulas (27) (or in the case  $U\rightarrow \infty$ approximation (28)) with (29) allows us to self-consistently compute $n_q$ and $G_p(z)$. 
The last thing needed for using this procedure is to rewrite formulas (22,24) in integral form using the definition of the coupling function 
$\Gamma_{\alpha m}(z)=2\pi\sum_k |t_{\alpha m}|^2\delta(z-\epsilon_{\alpha k m})$. For example

\begin{align}
\Sigma_p^{(0)}(z)=\sum_{\alpha k}\frac{|t_{\alpha p}|^2}{z-\epsilon_{\alpha k p}}=
\frac{1}{2\pi} \int \frac{\Gamma_{\alpha p}(\xi) d\xi}{z-\xi}\;.
\end{align}

Therefore by choosing a reasonable predefined form of coupling function  $\Gamma_{\alpha m}(z)$
we can self-consistency solve equation (28,29) and find the retarded Green function 
$G_p(z+i\eta)$ by using complex energy argument with sufficiently small imaginary part $\eta \rightarrow 0^{+}$
and further use it for calculating the current by formula (7).

\section{Numerical results}
We now apply the obtained formulas for nonequilibrium transport in nanotube quantum dot with the central part coupled to interacting
nanotube leads. In our numerical calculations we assumed that nanotube quantum dot is 
symmetrically coupled to contacts from nanotubes with Lorentzian linewidth $2D$, $\Gamma_{Lp}(z)=\Gamma_{Rp}(z)=\frac{\gamma D^2}{D^2+z^2}$.
From formula (30) it is obvious that all energy scale is dictated by the coupling function $\gamma$. 
In other words setting  $\gamma=1$  means that the energy is measured in 
$\gamma$ units. For comparison with experiments and other theoretical works the resulted quantities can 
be transformed in electronVolt energy scale by adopting 
$\gamma=10meV$~\cite{coupling}. For example the energy cutoff parameters $D=500$ used in our computation means $D=500\gamma=5 eV$. The dimensionless temperature 
$T=2$ corresponds to  $T=2\gamma \frac{K}{eV}=20 mK$. The contact dot interaction we used a flat-band profile 
$I_{\alpha}(z)=i_{\alpha}\theta(D-|z|)$, where $i_L,i_R$ are parameters that describe Coulomb interaction with left and right leads respectively. 
We next chose parameters which concerns the nanotube dot region.
 For nanotube quantum dot we adopted the dependence 
$\varepsilon_m=\varepsilon_{\{\sigma,\lambda\}}=\varepsilon_d-\sigma \lambda \Delta_{SO}/2$, here $\varepsilon_d$ is the basic dot level.
The value of  $\varepsilon_d$ was taken from our previous calculation~\cite{one} which accounts for the geometry of the NTQD $(5,5)/(10,0)\_l/(5,5)$ and depends on the length $l$ (here $l=1$ one unit cell) of the central fragment. By varying the length, different set of peaks appear in the density of states of the fragment. The tight binding approach used in this work did not include accounts electron-electron interactions. The present work does so, but it does not include the actual atomic structure of the nanotube system. We only took one peak closest to the Fermi level and set  $\varepsilon_d$ to its energy. For $(5,5)/(10,0)\_1/(5,5)$ nanotube quantum dot  $\varepsilon_d=0.006 t_{C-C}$ was calculated by us ~\cite{one}. Using for graphene hopping parameter the value~\cite{ten} $t_{C-C}\approx 2.66 eV$ we get in dimensionless units  $\varepsilon_d=15meV=1.5\gamma$. The model used does not either consider spatial spin correlation which is taken into account through some sort of mean field approximation through $\Delta_{SO}$.
in which the main contribution gives curvature~\cite{eleven} coupling $\Delta_{SO}^{curv}\approx 1.6 meV / d[nm]$.
Diameter of (10,0) nanotube is $d=3.9 nm$ which gives  $\Delta_{SO}=0.2 meV=0.02\gamma$. 

Applying spin bias $V_s$ means $\mu_{L\uparrow \lambda}=-\mu_{L\downarrow \lambda}=V_s/2$ and $\mu_{R\uparrow \lambda}=-\mu_{R\downarrow \lambda}=-V_s/2$. 
By swap indexes L with R or which is the same assigning 
 $\mu_{L\uparrow \lambda}=-\mu_{L\downarrow \lambda}=-V_s/2$, $\mu_{R\uparrow \lambda}=-\mu_{R\downarrow \lambda}=V_s/2$. 
 It is possible to mirror this dependence by OY axis but we decided to leave it unchanged for reason of distinguishing between 
 graphs for applied spin and conventional bias on basis of slope of a such dependences.

\myfig{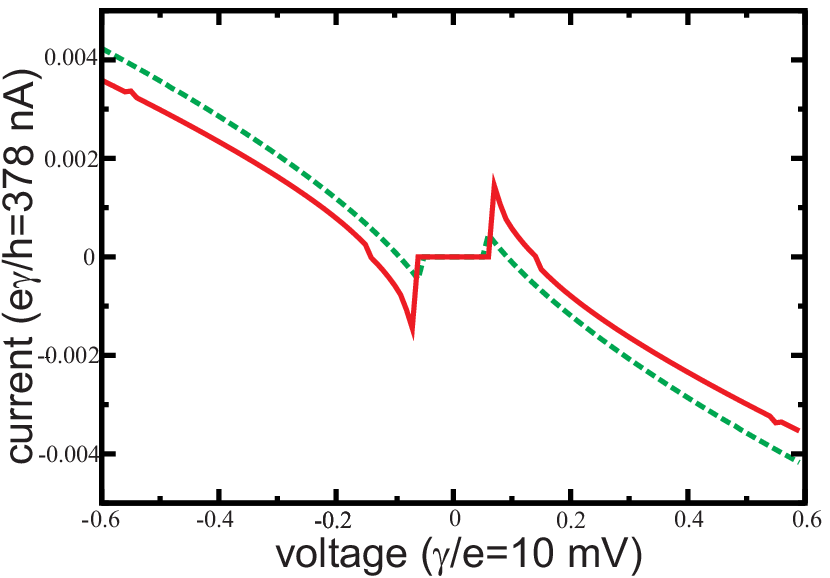}{Current for the case of spin bias voltage. Two states $m=\{\sigma\}$ QD at temperature T=1 without QD-lead interaction $i_L=i_R=0$   (green line) and with $i_L=i_R=0.05$ (red line).}{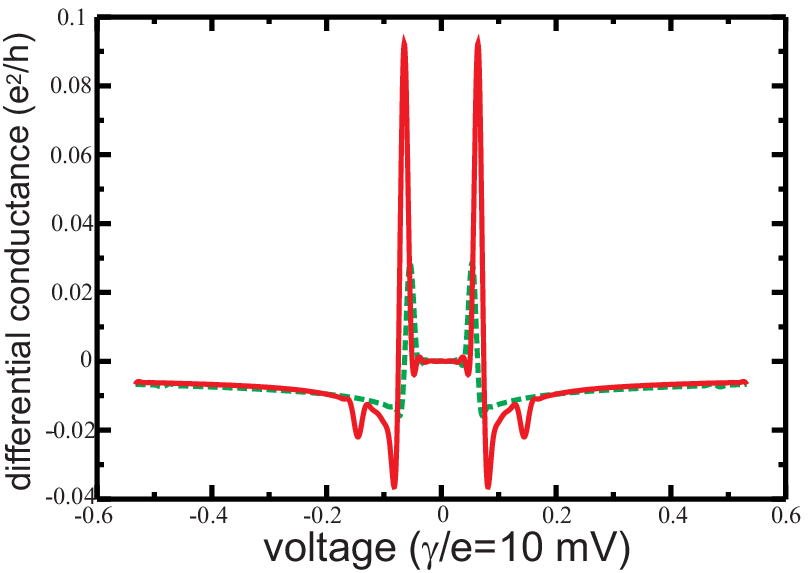}{Conductance for the case of spin bias voltage. Two states $m=\{\sigma\}$ QD at temperature  $T=1$ without QD-lead interaction $i_L=i_R=0$  (green) and with $i_L=i_R=0.05$ (red line)}

In this work we used $U\rightarrow \infty$ approximation. Studying the effect of $U$ will be devoted to another work. First (see Fig. 1) 
 we calculated the current-voltage characteristics for spin bias at $T=1$ temperature for different Coulomb coupling to contacts
 $i_L=i_R=0$ � green,  $i_L=i_R=0.05$ red curve. We also plotted its derivative (see Fig. 2), the
 so called differential conductance, as a function of applied spin bias $V_s$. 
 In all graphs dimensionless quantities are used. To get dimension quantities we need to take $\gamma$ value (for example 
 $\gamma=10 meV$). The position of peaks on Fig 1. near 0.1 in dimensionless units means 
 $0.1*10 \frac{meV}{e}=1mV$. On the Current-Voltage characteristics there is a gap which 
 is connected with Coulomb blockade regime. As we can see from Fig. 3 calculated for higher temperature 
 $T=100$, at this temperature the gap is already washed out. At low temperatures the differential conductance (Fig. 2) exhibits 
 narrow peaks and the nearest to zero voltage peaks are sufficiently extended in the case of presence of Coulomb 
 interactions with leads  $i_L=i_R=0.05$.

\myfig{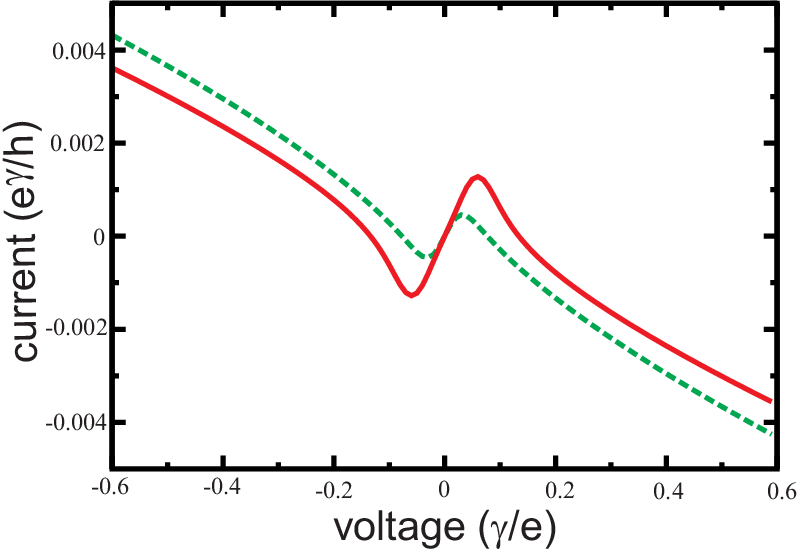}{Current for the case of spin bias voltage. Two states $m=\{\sigma\}$ QD at temperature $T=100$ without QD-lead interaction  $i_L=i_R=0$  (green) and with  $i_L=i_R=0.05$ (red)}{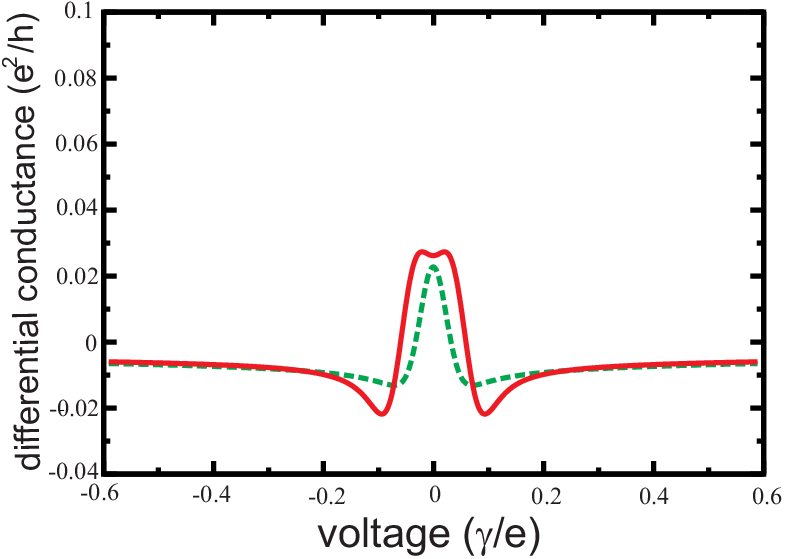}{Conductance for the case of spin bias voltage. Two states $m=\{\sigma\}$ at temperature  $T=100$ without QD-lead interaction $i_L=i_R=0$ (green) and with $i_L=i_R=0.05$ (red)}
\myfig{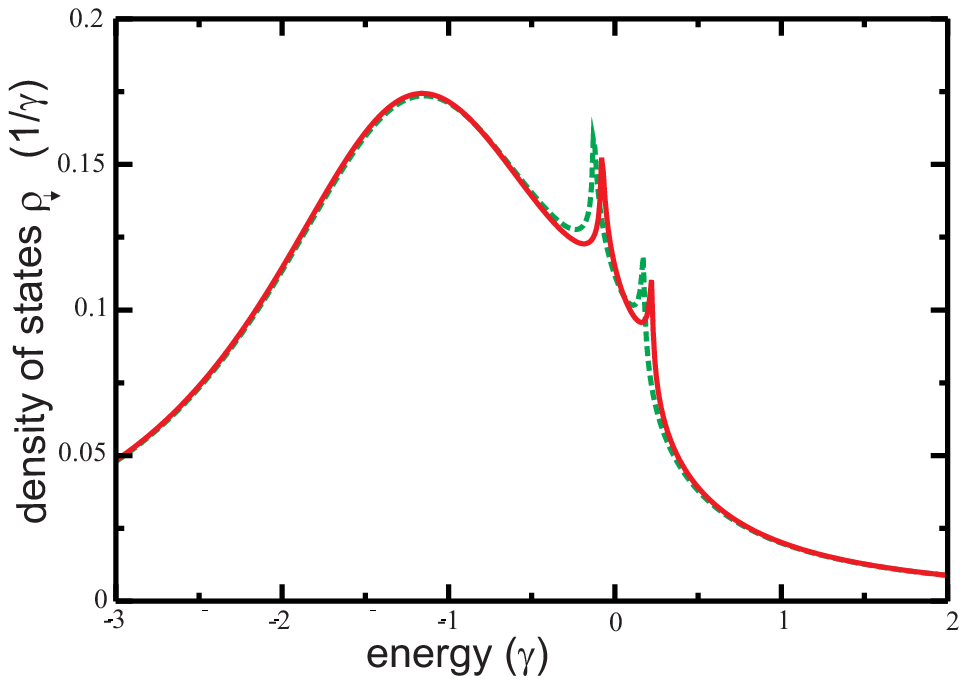}{Density of states for QD with two quantum states $m=\{\sigma\}$, a spin voltage of  $v_s=0.3$, spin down as well as without (green line) and with (red line) a QD-lead Coulombic interaction }{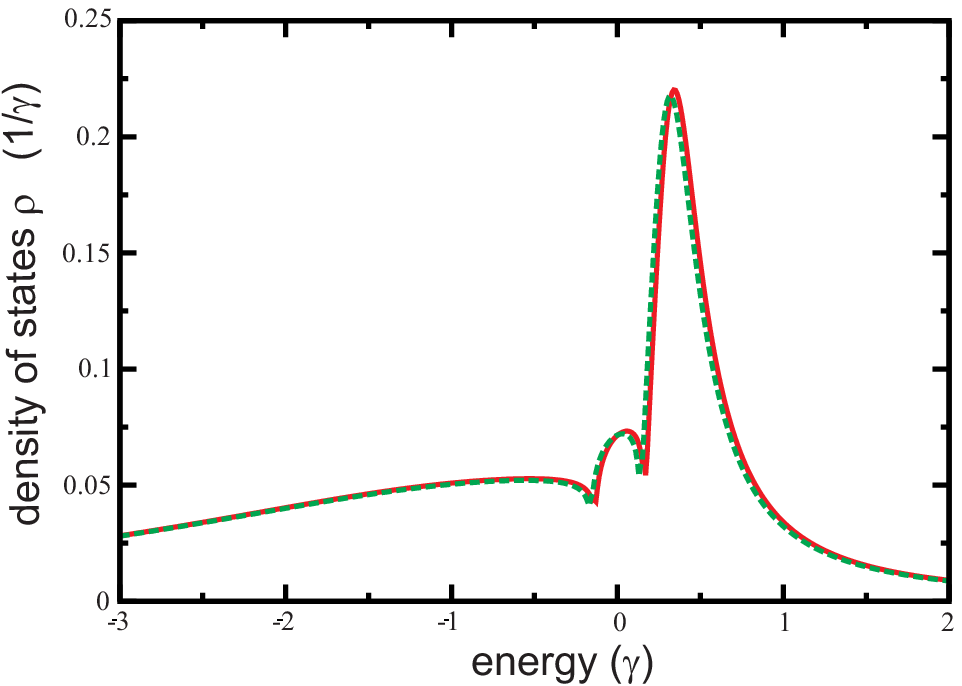}{Density of states for QD with four quantum states $m=\{\sigma, \lambda\}$ and a spin voltage of $V_s=0.3$ when $m=\{+,+\}$ or $\{-,-\}$  green) and $m=\{+,-\}$ or $m=\{-,+\}$ (red line).}
\myfig{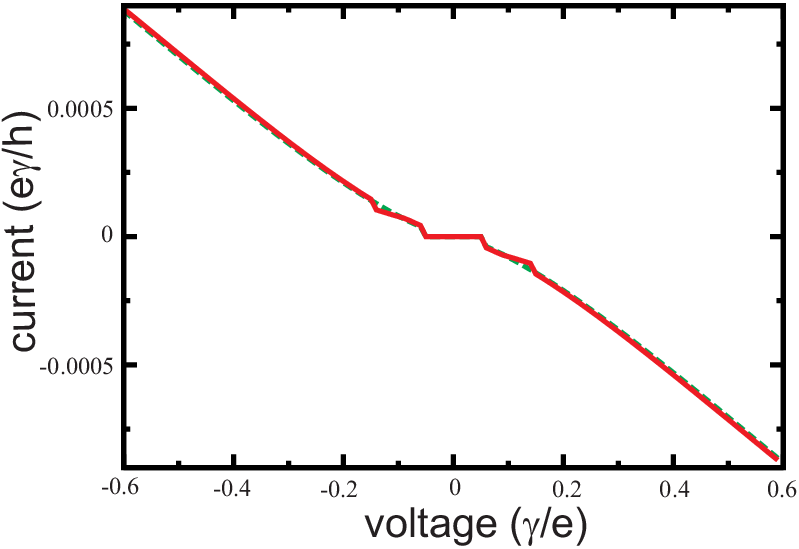}{Current for the case of spin-bias voltage.  Four states $m=\{\sigma, \lambda\}$  QD at temperature T=1, without QD-lead interaction   $i_L=i_R=0$ (green) and with $i_L=i_R=0.05$ (red line) }{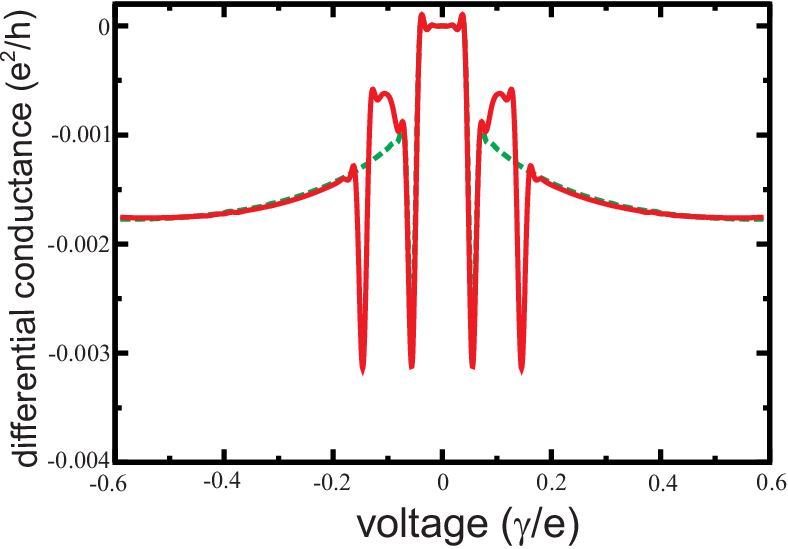}{ Conductance for the case of spin-bias voltage. Four states $m=\{\sigma, \lambda\}$ QD at temperature T=1, without QD-lead interaction  $i_L=i_R=0$ (green) and with $i_L=i_R=0.05$ (red line)}
\myfig{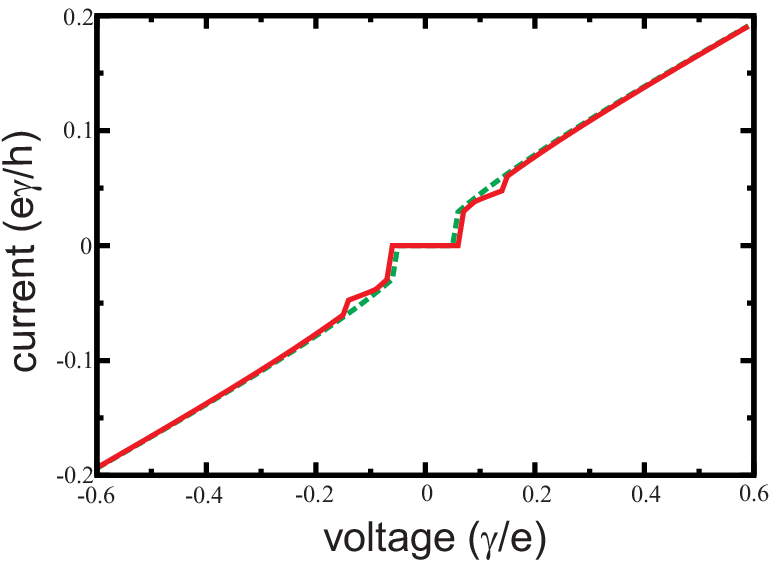}{Current for the case of conventional voltage. Two states $m=\{\sigma\}$  QD at temperature T=1  without QD-lead interaction $i_L=i_R=0$ (green) and with $i_L=i_R=0.05$ (red line)}{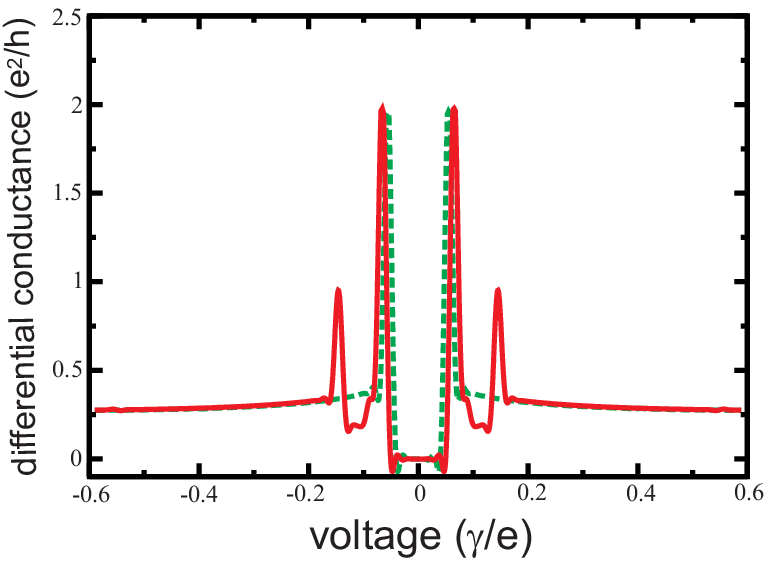}{ Conductance for the case of conventional voltage. Two states $m=\{\sigma\}$  QD at temperature T=1  without QD-lead interaction $i_L=i_R=0$ (green) and with $i_L=i_R=0.05$ (red line) }
\myfig{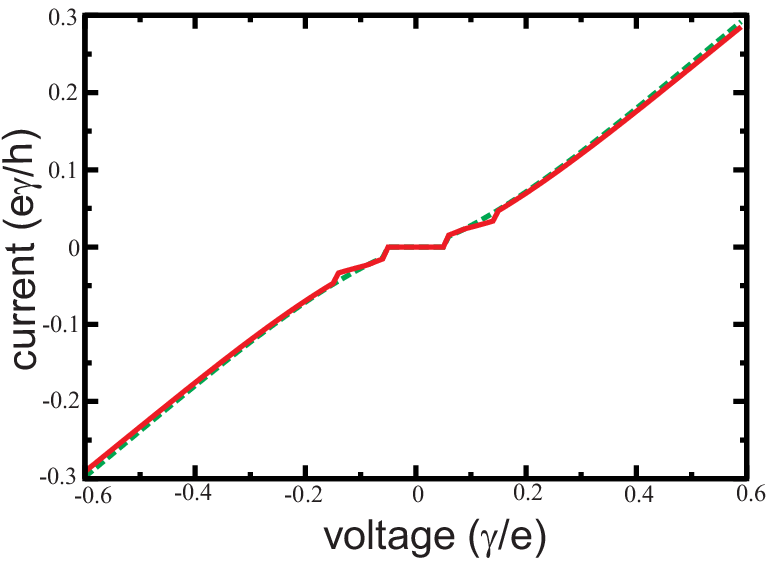}{ Current for the case of conventional voltage. Four states $m=\{\sigma, \lambda\}$ QD at temperature T=1, without QD-lead interaction $i_L=i_R=0$ (green) and with $i_L=i_R=0.05$ (red line) }{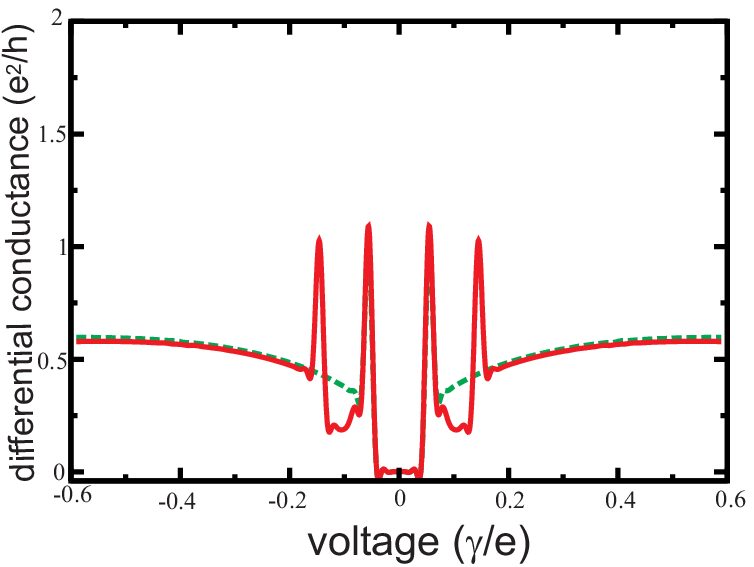}{Conductance for the case of conventional voltage. Four states $m=\{\sigma, \lambda\}$  QD at temperature T=1, without QD-lead interaction $i_L=i_R=0$ (green) and with $i_L=i_R=0.05$ (red line)}

 As we can see on Fig. 3,4 the temperature increase blurs the peaks of the differential conductance and smooth the current-voltage dependence.
 Therefore the interplay between the Kondo effect and the bias is highly temperature-dependent and is significant only at low temperature.

We also analyzed the density of states of quantum dot $\rho_p(z)=-(1/\pi)Im G_p^r(z)$ at applied spin-bias voltage $V_s=0.3$.
Fig. 5 depicts the calculated density of states for spin down electrons for different Coulomb QD-leads interaction.
The absence of interaction ($i_L=i_R=0$) is depicted by green curve, the presence ($i_L=i_R=0.05$) plotted by the  red curve. In both cases voltage 
$V_s=0.3$   is applied. As we can see there are two Kondo peaks which are split from each other due to the applied Voltage. 
When $V_s=0$ our calculation gives one Kondo peak. These results are in good agreement with~\cite{two}.

As we can see from Fig. 5, the Coulomb dot-lead  interaction shifts the density of states in high energy region. 
As we can see the position of nearest to zero conductance peaks are not affected by the strength of the Coulomb dot-lead  interaction.

The next step was to study the nanotube QD in which four states $m=\{\sigma, \lambda\}=\{\pm,\pm\}$ exist. In this case it is necessary to sum from q=1 to 4 in formulas like (28).
As earlier we consider spin bias voltage. Fig. 6 depicts calculated density of states for four different quantum states
$m=\{\sigma, \lambda\}=\{\pm,\pm\}$. As we can see the existence of four quantum states 
leads to abrupt changes in the density of states (see Fig. 5,6). In Fig. 7 the current-voltage characteristic for applied spin-bias voltage is presented.
As we can see this dependences is about 10 times less than that for the case of  two states 
 $m=\{\sigma\}=\{\pm\}$ in quantum dot. The second main difference in differential conductance is the presence of a second pair of strong peaks in the case 
 $i_L=i_R=0.05$ which is absent in the case $i_L=i_R=0$.

 Lastly, we investigated the effect of conventional bias, namely $\mu_{L\uparrow \lambda}=\mu_{L\downarrow \lambda}=-V/2$
 and $\mu_{R\uparrow \lambda}=\mu_{R\downarrow \lambda}=V/2$. The same parameters (Fig. 1,2) as in the case of conventional bias 
 (Fig. 9,10) give about hundred times lager current and lead to a different structure of peaks of conductance.

The conventional  bias in the case of four states  $m=\{\sigma, \lambda\}=\{\pm,\pm\}$ (nanotube quantum dot) 
gives (Fig. 11,12) about three hundred times lager current and yields a different structure of peaks  in the differential
conductance.  As we can see (Fig. 8,10,12) the presence of QD-lead interaction leads to formation of new pair of peaks.
The present results allows us to conclude that the QD-lead interaction is important and in all cases where it is strong it must be considered. 

\section{Conclusions}

We derived generalized Kubo formula for the computation of conductivity NTQD and formulas for self-consistent evaluation Green's function of NTQD in presence of Coulomb repulsion of electrons in QD region and electrons in lead. We show that presence of four states in NTQD with  m = \{$\sigma$, $\lambda$\} = \{$\pm$,$ \pm$\} leads to abrupt changes in the density of electronic states and to lowering the current amplitude by  approximately a factor of ten compared to  QD with only two states m = \{$\sigma$\} = \{$\pm$\} (without pseudospin). But in case of conventional bias, current amplitudes are approximately the same. We also show that the Coulomb interaction $i_L$ and $i_R$ with the left and right leads shifts the QD electron density of states to high energy region and the interplay between the Kondo effect and the bias is highly temperature-dependent and becomes significant only at low temperatures. Taking into account Coulomb dot-lead interaction leads to the emergence of a second pair of peaks in the differential conductance with significant amplitude. We also show that a such an interaction does not shift the  pair of peaks  (first pair) nearest to the Fermi level.

\section{Acknowledgments}

We thank Philippe Lambin for discussions. 

\bibliography{ogloblya}
\end{document}